# Direct observation of three-dimensional atomic structure of twinned metallic nanoparticles and their catalytic properties

Juhyeok Lee[1], Chaehwa Jeong[1] and Yongsoo Yang[1*]

[1]Department of Physics, Korea Advanced Institute of Science and Technology (KAIST), Daejeon 34141, Korea

**Abstract**

We determined a full 3D atomic structure of a dumbbell-shaped Pt nanoparticle formed by a coalescence of two nanoclusters using deep learning assisted atomic electron tomography. Formation of double twin boundary was clearly observed at the interface, while substantial anisotropy and disorder were also found throughout the nanodumbbell. This suggests that the diffusion of interfacial atoms mainly governed the coalescence process, but other dynamic processes such as surface restructuring and plastic deformation were also involved. A full 3D strain tensor was clearly mapped, which allows direct calculation of the oxygen reduction reaction activity at the surface. Strong tensile strain was found at the protruded region of the nanodumbbell, which results in an improved catalytic activity on {100} facets. This work provides important clues regarding the coalescence mechanism and the relation between the atomic structure and catalytic property at the single-atom level.

# Introduction

Nanoparticles have drawn considerable attraction for their broad applications in many scientific fields including physics, chemistry, medicine, materials sciences, and others[1–8]. Especially, metallic nanoparticles have shown great potential in the sense that their catalytic performance can be finely controlled based on growth conditions. Substantial efforts have been devoted to realizing their desired functional properties by controlling specific size, shape, and composition of the synthesized nanocrystals[9–14]. Recent studies have revealed that the classical monomer attachment process is not the only mechanism for nanocrystal growth, and that particle-particle interaction, i.e., coalescence, also plays a crucial role in determining the size, shape, and structure of the nanocrystals[15–17]. Therefore, the coalescence mechanism has been actively studied by theoretical calculations[18–25] as well as in-situ measurements[9,11,12,16,17,26,27], and it has been shown that a coalescence event can result in not only single-crystalline nanoparticles, but also (multiply) twinned structures[11,17,26,28–31]. Since partial coalescence can also be beneficial for catalytic performance[32–35], a better understanding of the coalescence and twinning process is important for fine tailoring of the nanocrystal structure and related catalytic properties. However, most of the experimental studies have been limited to 2D observations. Nanocrystals grow in a 3D space, and to truly understand the twinning mechanism, it is essential to study the interface between nanoclusters in 3D atomic details. Moreover, the 2D measurements are mostly limited to structural analysis, and the information contained in 2D projections is not enough to be directly related to material properties. On the other hand, 3D atomic structural information can not only provide the atomic arrangements at twinned interfaces, but also directly yields the 3D strain as well as material properties via quantum mechanical calculations such as density functional theory (DFT). For example, 3D surface local strain obtained from the 3D atomic structure can be linked to a catalytic performance like oxygen reduction reaction (ORR)[7,36]. Therefore, it is necessary to go beyond the 2D measurements and precisely determine the 3D atomic structures of the partially-merged nanocrystals to fully understand the twinned interface structure and resulting physical properties.

For the 3D structural analysis, atomic electron tomography (AET) can be a powerful tool which allows direct determination of 3D atomic structures without assuming crystallinity[37–42]. Rich structural information, including point defects/dislocations[38,39,43], chemical order/disorder[38], nucleation[44], and 3D strain[39,43], has been revealed at an atomic resolution. Recently developed deep learning based neural network-assisted AET resolved the artifacts from the data imperfection and further enhanced the reliability of the determined atomic structures[45].

Here, we applied the neural network assisted AET to determine a 3D atomic structure of a dumbbell-shaped Platinum (Pt) nanoparticle. A doubly twinned structure with a stacking fault was clearly observed at the interface, together with overall anisotropy and disorder, suggesting a pathway of coalescence process expected from MD simulations. Furthermore, the 3D atomic structure was directly related to catalytic properties via surface strain analysis, proving that the strain induced from the coalescence process can substantially affect the facet-dependent ORR.

## Results

### Determination of 3D atomic coordinates

A Pt nanoparticle specimen was prepared by drop-casting onto a carbon membrane followed by a vacuum annealing at 120 °C for 24 hours (see Methods in Supporting Information). A tomographic tilt series of 21 projections with the tilt angles ranging from -70.0° to +70.0° was acquired for a dumbbell-shaped Pt nanoparticle using an aberration-corrected transmission electron microscope operated in annular dark-field

(ADF) mode (see Methods in Supporting Information). After image-post processing (image de-noising and alignment), a 3D tomogram was reconstructed from the tilt series using GENFIRE algorithm[46] (see Methods in Supporting Information). The tomogram was further augmented by a deep learning based neural network to correct for artifacts which can arise from data imperfection due to sparse sampling and missing wedge problem[45]. Atom-tracing from the 3D volume revealed that the nanodumbbell contained 3504 Pt atoms in total (see Methods in Supporting Information).

To evaluate the goodness of the determined atomic structure, we calculated the R-factors between the simulated tilt series forward-projected from the determined atomic structure and the experimentally acquired tilt series (see Methods in Supporting Information). The averaged R-factor was 0.158 before applying the deep learning augmentation, which was improved to 0.151 after applying the deep learning augmentation. We also performed a precision analysis using multislice simulation to estimate the validity of the 3D reconstruction, which showed 0.23% error of atom identification and a 3D precision of 9.9 pm (Fig. S1, see Methods in Supporting Information).

To further ensure the reliability of our deep learning based approach, we checked the Fourier projections of the 3D tomograms (Fig. S2) and the averaged atomic profile (Fig. S3), which show that the missing information has been successfully retrieved by the deep learning based augmentation. We also tested our result using a deep learning neural network trained by amorphous structural models rather than those of face-centered cubic (f.c.c.). It can be clearly seen that the use of the neural network trained by amorphous models also provides a consistently improved 3D tomogram and resulting atomic structure (see Methods in Supporting Information). The atomic structure obtained from the tomogram augmented with the neural network trained by f.c.c. based models was chosen for further analysis since it showed the smallest R-factor (see Methods in Supporting Information).

**Structural analysis**

Figure 1 shows the experimentally determined atomic structure of the Pt nanodumbbell and its forward-projection images along different crystallographic directions. As shown in Fig. 1a-e, the nanocrystal is clearly showing a dumbbell shape, suggesting that it has been formed by coalescence of two nanoclusters. Note that no sign of twin boundaries can be found in three of the forward projection images (Fig. 1f-h), which are along the $z$ direction in the lab-coordinates (i.e., that for the 0° image of the tilt series), along the $[1\bar{2}1]$ direction, and along the $[\bar{1}\bar{1}1]$ direction, respectively. If we acquire 2D measurements only along these directions, we may conclude that this nanodumbbell is a single crystal. Interestingly, the forward projection image along the [001] direction shows some blurred contrast at the interface, indicating that this nanocrystal is not a single crystal and there can be an interesting interface structure (Fig. 1i). However, it is difficult to obtain more detailed structural information from this image only. The projection image along the [1$\bar{1}$0] direction, on the other hand, is showing a clear double twin boundary structure (Fig. 1j). This shows that 2D measurements are not enough to clearly reveal the 3D interface structure, and they can even be misleading, emphasizing again that it is essential to have the 3D structural information to fully understand the twinned interface structure.

The 3D atomic structure clearly shows that the two nanoclusters share common (111) faces at the interface (Fig. 1e,j and Fig. 2a) with a stacking fault (the atomic layer with the red atoms in Fig. 2a). As shown in Fig. 2b, each nanocluster forming the nanodumbbell exhibits a f.c.c. structure, which can be represented as the atomic layers of 'ABC' stacking order perpendicular to the [111] direction. However, a clear stacking fault was found at the interface between the two nanoclusters. The red atomic layer in Fig. 2a, which has 'A' type of atomic arrangement, exists between 'B' and 'C' type layers (yellow and green atomic layers in Fig. 2a, respectively) of otherwise perfect f.c.c. ordered layers. This directly results in two locally h.c.p. ordered layers (twin boundaries), which can be represented as 'ABA' stacking order or 'ACA' stacking order, as shown in Fig. 2a,c-g. Note that the same stacking fault was clearly observed from the tomogram augmented with the neural network trained by amorphous structural models (Fig. S4), evidencing that the observed twinned structure is genuine and not from deep learning artifact.

**Strain analysis**

From the 3D atomic structure of the dumbbell-shaped Pt nanoparticle, we calculated the 3D strain tensors for the two nanoclusters separately (Fig. 3a) (see Methods in Supporting Information). Figure 3b-g show the six components of the strain tensor. In overall, strong anisotropy can be observed from the strain map. The bottom half of the nanodumbbell exhibits strong compressive and tensile $\varepsilon_{xx}$ strains alternatively distributed on different faces (Fig. 3b), while $\varepsilon_{yy}$ strains show relatively weak but similarly anisotropic behavior (Fig. 3c). For the top half of the nanodumbbell, on the other hand, strong tensile $\varepsilon_{xx}$ strains dominate especially along the $y$ ([$\bar{1}$10]) direction, with $\varepsilon_{yy}$ strains being mainly compressive (Fig. 3b-c). Unlike other principal strains, the magnitude of most $\varepsilon_{zz}$ strains are less than 1% (Fig. 3d), suggesting that there is no strong disorder along the $z$ ([111]) direction unlike the observed strong anisotropic disorders along the $x$ and $y$ directions. Interestingly, there is a clear protruded region in the top nanocluster (Fig. S5), which coincides with the region of the strong tensile $\varepsilon_{xx}$ strains and compressive $\varepsilon_{yy}$ strains. The overall anisotropic behavior and the strong strain near the protruded region are likely to be deeply related to the observed disorders resulting from coalescence dynamics.

**Suggested coalescence pathway**

It can be expected that our observed doubly twinned interface structure (Fig. 1j and Fig. 2a,c,d) is formed by a diffusion process of interfacial atoms, because the doubly twinned structure cannot be formed by simply attaching two f.c.c. nanoclusters. Moreover, the double twin boundary structure over three-atomic layers were already predicted by a molecular dynamics (MD) simulation[24]. The simulation also suggests that the double twin boundary can be formed via diffusion restructuring process when two nanoclusters are well aligned to each other with common (111) faces without an angular misorientation. Considering the similarity between our experimentally observed interface structure sharing (111) faces and that from the MD simulation, it is natural to assume that the doubly twinned interface was formed mainly by diffusion process starting from a well-aligned initial condition.

However, the detailed structural behavior of the nanodumbbell cannot be explained using a simple diffusion-based merging process of well-aligned nanoclusters. We observed several quantitative evidences suggesting that surface restructuring and misorientation-induced dynamic processes (such as plastic deformation) also have been involved during the formation of the nanodumbbell.

First of all, the atomic layers near the interface show much larger disorder compared to the other layers. We compared each atomic layer (perpendicular to [111] direction) with an ideal 2D hexagonal lattice and calculated the 2D root-mean-square deviation (RMSD) for each layer. It can be clearly seen that the interfacial atomic layer and the adjacent twin boundary atomic layers have higher RMSD values than that of the other atomic layers (Fig. S6b), indicating larger disorder. For cross-checking of the disorder estimation, we also calculated the bond orientation order (BOO) parameters of the Pt nanodumbbell (see Methods in Supporting Information). The $Q_4$ and $Q_6$ maps in Fig. S6c show that the interface and the twin boundaries are the most disordered, which is consistent with the results from the 2D RMSD calculation. Second, surface atoms exhibit higher 2D RMSDs than those of core atoms in all atomic layers perpendicular to the [111] direction (Fig. S6b). This feature can also be clearly seen from Fig. 2e-g; the surface atoms are deviating from the ideal hexagonal lattice more severely. Third, there is a protruded region in the top nanocluster (Fig. S5), which clearly shows higher strain (Fig. 3) and higher disorder (i.e., larger 2D RMSD, see Fig. S6b) compared to other regions. Fourth, there is slight relative rotation (1.6°) between the top and bottom nanoclusters along [111] direction (see Methods in Supporting Information).

The anisotropy and structural disorders described above suggest that the merging process of two nanooclusters was accompanied by complicated coalescence dynamics as well as the diffusion process. Considering the overall twisting (relative rotation between the top and bottom nanoclusters), it is expected that there was slight angular misorientation between the two nanoclusters right before the coalescence event. Then, during the coalescence, connection between the nanoclusters (neck formation) can induce rotational restriction, resulting strong internal stress, which can form the protruded region in the merged nanocrystal with a strong tensile strain[21,25]. The large disorder of both surface and interface suggests that further stabilization processes such as surface restructuring and plastic deformation were involved during the coalescence. To fully understand the suggested coalescence pathway and elucidate the exact coalescence mechanisms, more dynamical, theoretical, and simulational studies are necessary.

**ORR calculation**

The observed anisotropy and disorders induced by the coalescence event will result in strong anisotropic surface strains, which can critically affect the catalytic performance. Since we have the full 3D surface atomic structure, it can be directly correlated to the catalytic properties to reveal the strain-induced effect. To evaluate the surface ORR electrocatalytic activity of our measured Pt nanodumbbell, we first quantified the relations between the surface strain, OH binding energy, and ORR activity. The relation between the ORR activity and OH binding energy for Pt (111) and (100) model surfaces was reported in previous studies[47,48] using the microkinetic model (Fig. S7a). Another relation between the OH binding energy and the strain at Pt (111) and (100) model surfaces can also be obtained via density function theory calculations[49] (Fig. S7b). From the two reported relations, the surface strain and ORR activity of Pt (111) and (100) surfaces can be directly correlated by volcano-shaped curves (Fig. S7c). Next, we calculated the surface local strain for each atom on {111} and {100} facets from the experimentally determined 3D atomic structure (see Methods in Supporting Information). Figure 4a-f show the obtained surface local strain maps on {111} facets, {100} facets, and both facets. Note that the surface local strain of the protruded region (Fig. S5) shows strong tensile behavior (Fig. 4a,c,e), which is consistent with the strong tensile $\varepsilon_{xx}$ strain observed in the 3D strain map (Fig. 3b). Based on the calculated surface local strain and the volcano relation discussed above, full 3D maps for the ORR activity were obtained for {111} and {100} facets (Fig. 4g-l).

It is well-known that the ORR electrocatalytic activity of {111} facets is superior to that of {100} facets in the absence of surface strain (Fig. S7c). Our ORR results are in agreement with the general expectation (Fig. 4g-l). However, the regions of strong tensile and compressive strain show the opposite behavior (Fig. S7c). Figure 4a,c,g,i show that {100} facets can exhibit better ORR performance compared to that of {111} facets in the regions of strong tensile strain. The strong strain dependence of the ORR activity suggests that ORR can be finely tuned by controlling the surface strain. Considering the fact that the surface strain is sensitive to not only the coalescence pathway as discussed above but also surface boundary conditions[45] such as choice of support, surface coating, and core-shell design, the single-atom level ORR determination capability demonstrated here opens a door for the controlled design of nanocatalysts with desired properties via strain engineering.

## Conclusions

We determined the 3D atomic structure of the dumbbell-shaped Pt nanoparticle via neural network-assisted AET at the single-atom level. The measured 3D atomic structure revealed the twinned structure containing doubled twin boundaries at the interface between the two nanoclusters, together with strong anisotropy and disorders, suggesting a complex nanoparticle coalescence pathway based on multiple scenarios predicted by MD simulations. Strong anisotropic strain was clearly observed throughout the nanodumbbell as expected from the observed structural anisotropy and disorder, which is likely to be correlated with the internal stress induced during the coalescence event. Furthermore, the full 3D strain map and surface ORR activity were directly obtained from the atomic structure, showing that the ORR activities strongly depend on the facet structure as well as the surface strain. Looking forward, the capability of measuring 3D atomic structural details of twinned system as well as full 3D surface ORR map would deepen our understanding of the coalescence process and the catalytic structure-property relations, shedding light on the design and synthesis of fine-tailored nanocatalysts with desired properties.

# Figures

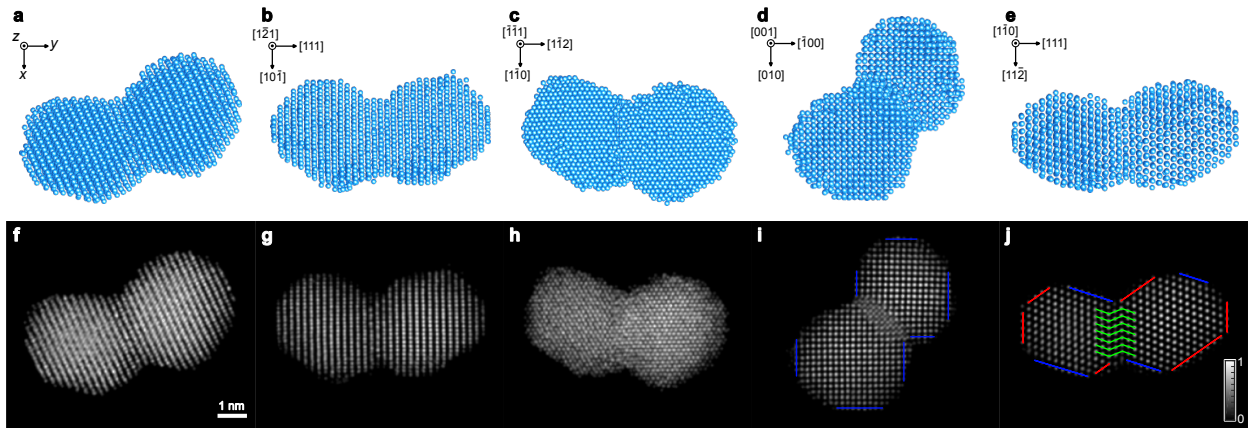

**Figure 1 | The experimentally measured 3D atomic structure of a Pt nanodumbbell and its forward-projection images along different crystallographic directions.** The 3D atomic structure visualized at different orientations, for which the paper is perpendicular to the $z$ direction of the experimental lab-coordinates (**a**), $[1\bar{2}1]$ direction (**b**), $[\bar{1}\bar{1}1]$ direction (**c**), $[001]$ direction (**d**), and $[1\bar{1}0]$ direction (**e**). The linear forward-projection images of the atomic structure along the $z$ direction of the lab-coordinates (**f**), $[1\bar{2}1]$ direction (**g**), $[\bar{1}\bar{1}1]$ direction (**h**), $[001]$ direction (**i**), and $[1\bar{1}0]$ direction (**j**). The red and blue lines in (**i**) and (**j**) represent {111} and {100} facets, respectively. The green lines in (**j**) are guides for the eye, highlighting the double twin boundaries.

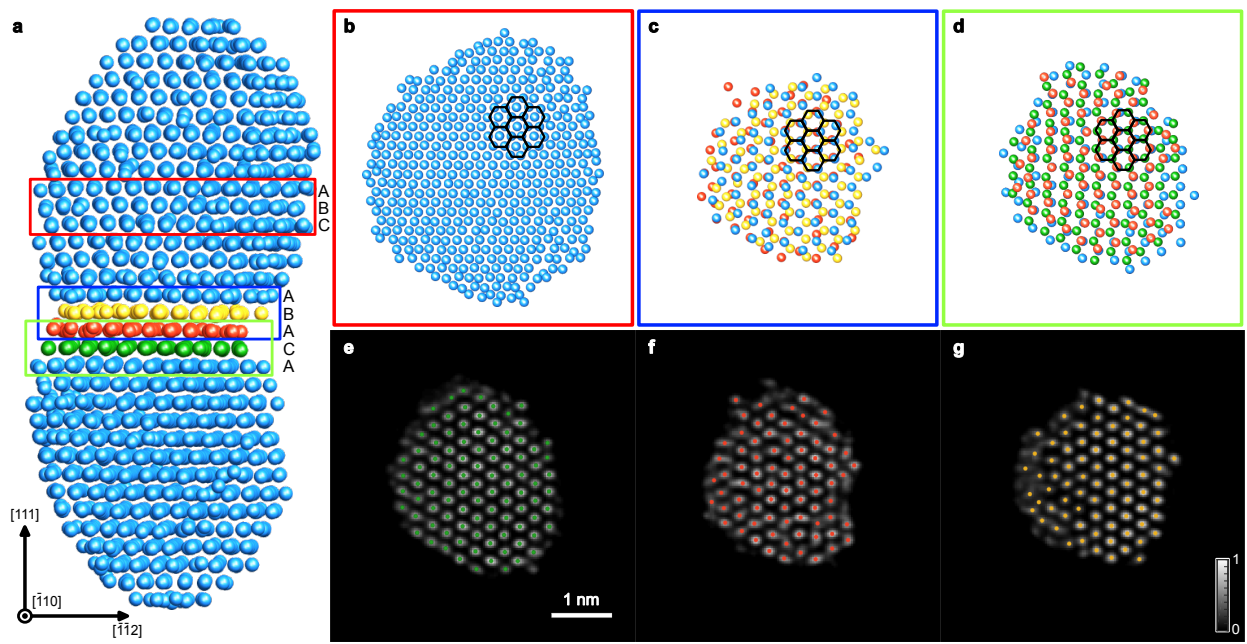

**Figure 2 | Overall 3D atomic structure, twin boundaries, and interface structure of the Pt nanodumbbell. a**. The 3D atomic structure which contains twin boundaries at the interface between the two nanoclusters. **b**. Three atomic layers, perpendicular to [111] direction [the atomic layers in the red box (**a**)], viewed along the [111] direction. **c**. Similar plot with (**b**), for the three atomic layers in the blue box in (**a**). **d**. Similar plot with (**b**), for the three atomic layers in the green box in (**a**). The layer of the red atoms represents the interface between the two nanoclusters, and the layers of yellow and green atoms are the upper and lower twin boundaries, respectively. The black hexagons in (**b-d**) are guidelines to highlight the f.c.c. and h.c.p. stacking orders. **e-g**. 2 Å thick slices of the 3D tomogram, perpendicular to the [111] direction (grayscale background) for the lower twin boundary (**e**), for the interface layer (**f**), and the upper twin boundary (**g**). Colored dots in (**e-g**) represent the traced atom positions.

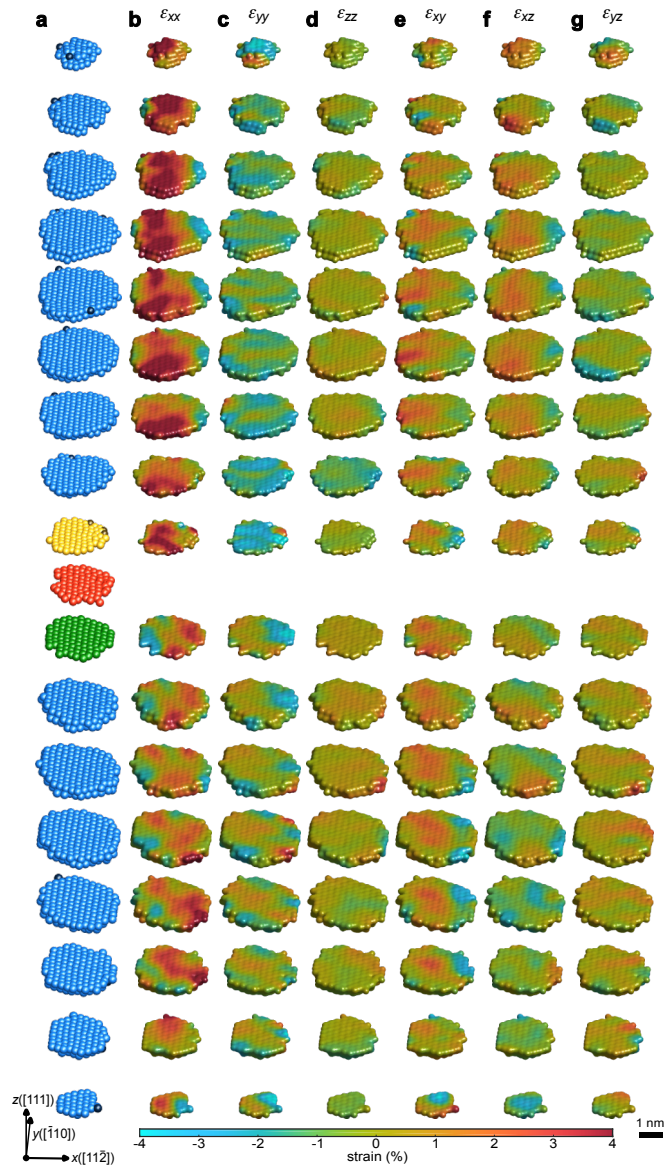

**Figure 3 | 3D Strain map of the Pt nanodumbbell. a**. Atomic structure of the Pt nanodumbbell sliced along the [111] direction for every two atomic layers, except for the interface and twin boundaries which are sliced into single atomic layers. The colors of the atoms follow the color convention in Fig. 2, except for the black atoms which are for the atoms not-assigned to f.c.c. lattice sites during the strain calculation process (see Methods in Supporting Information). The 3D strain maps represent $\varepsilon_{xx}$ (**b**), $\varepsilon_{yy}$ (**c**), $\varepsilon_{zz}$ (**d**), $\varepsilon_{xy}$ (**e**), $\varepsilon_{xz}$ (**f**), and $\varepsilon_{yz}$ (**g**) components of the strain tensor.

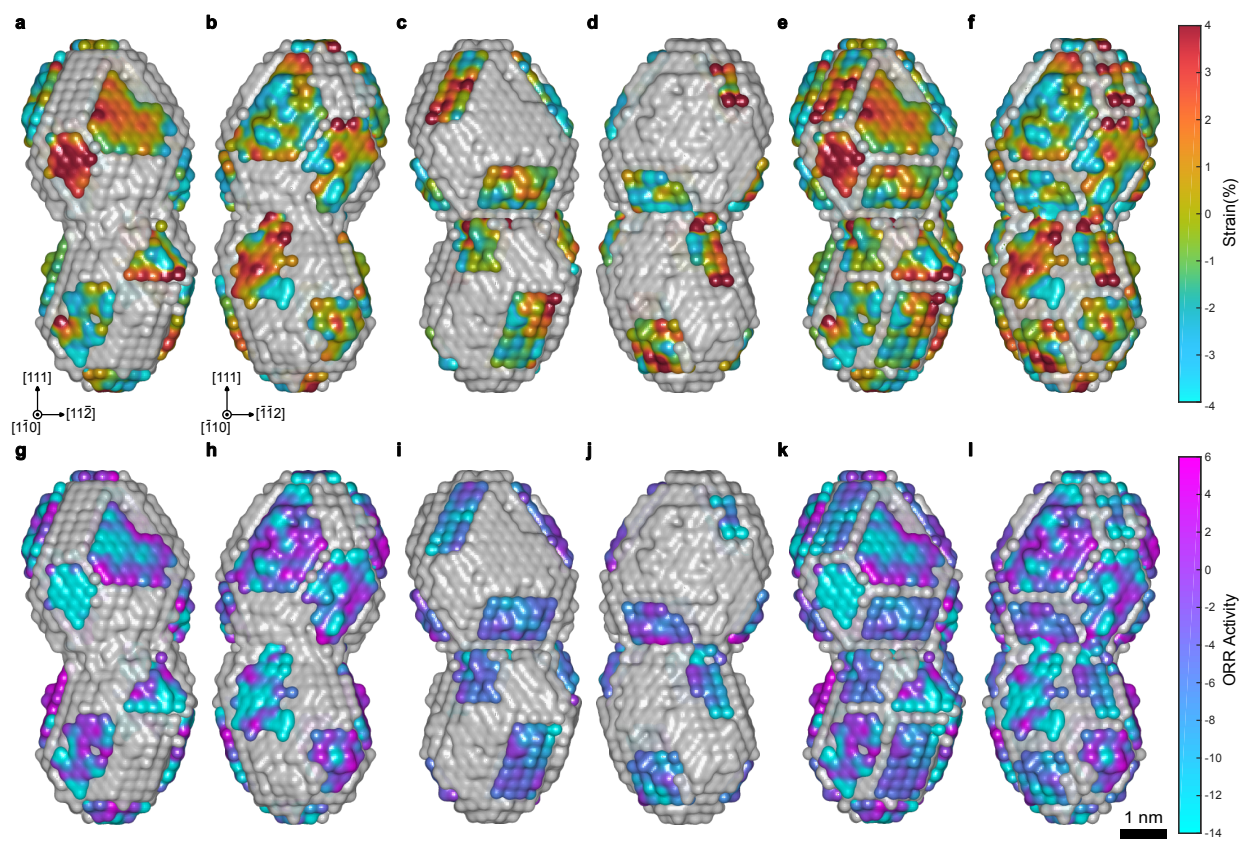

**Figure 4 | Surface local strain and ORR electrocatalytic activity of the Pt nanodumbbell on {111} and {100} facets. a-b**. Surface local strain of the Pt nanodumbbell on {111} facets viewed along two different orientations ([$\bar{1}$10] and [1$\bar{1}$0] directions for (**a**) and (**b**), respectively). **c-d**. Similar figures for {100} facets. **e-f**. Similar figures for both facets. **g-h**. Calculated ORR electrocatalytic activity of the Pt nanodumbbell on {111} facets viewed along two different orientations ([$\bar{1}$10] and [1$\bar{1}$0] directions for (**g**) and (**h**), respectively). **i-j**. Similar figures for {100} facets. **k-l**. Similar figures for both facets. The plotted ORR activity is represented as log($j/j_{Pt,(111)}$), where $j$ is the current density.

## Acknowledgements

We thank Elisabeth M. Dietze for helpful discussions. This research was supported by the National Research Foundation of Korea (NRF) Grants funded by the Korean Government (MSIT) (No. 2020R1C1C100623911). Y.Y. was partially supported by the KAIST singularity professor program. J.L. and C.J. were also partially supported by the KAIST-funded Global Singularity Research Program (M3I3) for 2019, 2020, and 2021. The STEM experiment was conducted using a double Cs corrected Titan cubed G2 60-300 (FEI) equipment at KAIST Analysis Center for Research Advancement (KARA). Excellent support by Hyung Bin Bae, Jin-Seok Choi, and the staff of KARA is gratefully acknowledged.

# Supporting Information

## for

Direct observation of three-dimensional atomic structure of twinned metallic nanoparticles and their catalytic properties

Juhyeok Lee[1], Chaehwa Jeong[1], and Yongsoo Yang[1*]

[1]*Department of Physics, Korea Advanced Institute of Science and Technology (KAIST), Daejeon 34141, Korea*

*Email: yongsoo.yang@kaist.ac.kr

## 1. Methods

**Data acquisition**

Pt nanoparticle solution supplied in aqueous 2 mM sodium citrate was purchased from nanoComposix. Pt specimen was prepared by drop-casting the solution onto a 4-nm thick carbon membrane, and was annealed in vacuum at 120 °C for 24 hours. A tomographic tilt series was acquired from a Pt nanodumbbell using a Titan double Cs corrected transmission electron microscope (Titan cubed G2 60-300). The images were acquired at 300 keV acceleration voltage in annular dark-field (ADF) mode with 25.1 mrad convergence beam semi-angle, 40 mrad and 200 mrad detector inner and outer semi-angles, respectively. A tilt series of 21 tilt angles ranging from -70.0° to 70.0° were collected. Three images per each tilt angle were measured with 4 μs dwell time, 13.1 pA beam current, and the scanning step size of 0.358 Å. The total electron dose for the entire tilt series was determined to be $1.61 \times 10^5$ e Å$^{-2}$. For checking the potential structural change resulting from electron beam during the measurement, we measured the zero-degree images before and after the experiment. We found that the Pt nanodumbbell was slightly rotated (Euler angles (z-y'-x'' convention) $\{\psi : -0.4°, \theta : 2.6°, \varphi : -3.8°\}$) during the experiment. The surface of the Pt nanodumbbell shows slight difference between pre- and post-measurement images, but the internal part (interface between the two nanoclusters) of the images are consistent. The consistency was further verified by forward-projecting the final atomic model and comparing the forward projection with the post-experiment zero-degree projection (see Fig. S8).

**Image post-processing and 3D reconstruction**

We performed the image post-processing (drift correction, scan distortion correction, BM3D denoising[1,2], background subtraction, and tilt-series alignment based on center-of-mass and common line method) for the raw tilt series of the Pt nanodumbbell following the procedures described in previous works[3–6]. We also applied a circle-shaped low-pass filter with a diameter of 2.08 Å$^{-1}$ (the boundary of the low-pass filter was slightly smoothened by 0.05 Å$^{-1}$) to the tilt series for reduction of unphysical high-frequency noise.

A 3D tomogram was reconstructed from the post-processed tilt series using GENFIRE algorithm[7]. To improve the quality of reconstruction, rotational and translational re-alignments were further applied. After the correction, the final 3D tomogram was computed using GENFIRE algorithm[7] with the following GENFIRE parameters: discrete Fourier transform (DFT) interpolation method, number of iterations 1,000, oversampling ratio of 5, and interpolation radius 0.1.

**Generation of input/target datasets for the deep-learning based neural network**

Input and target data were prepared for training, validation, and test of the neural networks following the previously reported procedures[8] as follows. A total of 24,000 (20,000 for training, 2,000 for validation, and 2,000 for test) simulated tomograms were generated for input and target data.

For the generation of input data, we followed a four-step process. First, we created 3D f.c.c. atomic structures with lattice constant 3.912 Å with random-shaped volumes ranging from 96,000 to 154,000 Å$^3$. Point defects (percentages of the point defects were randomly chosen between 0% and 0.5%) and random spatial displacement (approximately 22 pm RMSD) were added to the atomic structure. We also generated amorphous atomic structures. For generating amorphous structures, we randomly put atoms in random-shaped 3D volumes ranging from 96,000 to 154,000 Å$^3$ with the constraint of minimum distance 2.0 Å, targeting approximately 95% atom density compared to that of the f.c.c. structure. Second, the 3D atomic potentials[9] at the atomic positions were calculated and convolved with Gaussian kernels with standard deviation (σ) (the standard deviation (σ) was randomly chosen from a Gaussian distribution of mean 0.45 Å and standard deviation 0.11 Å) in the 3D volumes. Third, 21 forward-projections of the 3D volumes for tilt angles ranging from -65° to +65° were generated as a tilt series for each volume. The size of each projection was 256 × 256 pixels with pixel size 0.358 Å. To mimic the real experiment, Poisson noise and random tilt angle errors up to ±0.4° were added. Fourth, a 3D tomogram was computed from the generated tilt series and the tilt angles using the GENFIRE algorithm[7]. Fast Fourier transform (FFT) interpolation method, number of iterations 100, oversampling ratio of 2, and interpolation radius 0.3 were used for the GENFIRE reconstructions.

For the generation of target data (ground truth), we first followed the first step in the generating procedure of input data. Then, a 3D Gaussian distribution with standard deviation (σ) of 0.45 Å was placed at each atomic position in each 3D volume.

**Optimization of Gaussian broadening width for the training data**

To mimic the experimental electron beam size and thermal vibration for our input data, we applied a Gaussian kernel for the data generation process. The width of the Gaussian kernel was estimated as follows. We fitted a 3D Gaussian function of $11 \times 11 \times 11$ voxels to the averaged $11 \times 11 \times 11$ voxels extracted from the experimental 3D tomogram at every traced atom position. Among the three width parameters obtained from the 3D Gaussian fitting, we discarded one parameter which is along the missing wedge direction to prevent width overestimation. The geometrical mean of the remaining two fitted Gaussian widths was used as the optimized Gaussian width for training data preparation. The optimized Gaussian width was calculated to be 0.45 Å.

**The framework of deep learning neural network and training**

The basic structure of the deep learning neural network followed the framework suggested in the previous work[8]. The only major difference is the depth of the network; we used a neural network two layers deeper than that of the previous work[8] (see Fig. S9). The sizes of input and output data were both $256 \times 256 \times 256$ voxels. Main building blocks of the deep learning neural network are $3 \times 3 \times 3$ convolution with stride 2 for down-sampling, $2 \times 2 \times 2$ max-pooling for down-sampling, $3 \times 3 \times 3$ transposed convolution with stride 2 for up-sampling, and two $3 \times 3 \times 3$ convolution with stride 1. Dropout method[10] was also employed to prevent overfitting. Two activation functions (Leaky Rectified Linear Unit[11] with the coefficient of leakage 0.2 for all cases except for the last output, and Rectified Linear Unit[12] for the last output) were used. Mean-square error was used as the loss function, and the Adam optimizer[13] was used with the learning rate of $2 \times 10^{-4}$.

Figure S10 shows the learning curves for the training sets of the f.c.c. based model and the amorphous based model. There is no sign of overfitting or divergence in the learning curves. The final trained neural network was chosen at the 14th epoch for the training based on f.c.c. based structure and the 27th epoch for that of the amorphous based structure, where the loss function is clearly converged.

**Identification of 3D atomic coordinates**

To determine the 3D atomic coordinates, atom-tracing and atom classification methods described in previous works[4,5] were performed for 3D tomograms after applying the deep learning based neural networks. The numbers of traced atoms for the 3D tomograms after applying the deep learning based neural network trained by the f.c.c based atomic models and the amorphous based atomic models were 3,492 and 3,360, respectively. A manual correction was applied to add some atoms not properly traced during the automatic searching. For the traced atomic model from the 3D tomogram after applying the deep learning based neural network trained by the f.c.c. based atomic models, 12 Pt atoms (0.34%) were manually added. For the traced atomic model from the 3D tomogram after applying the deep learning based neural network trained by the amorphous based atomic models, 19 Pt atoms (0.56%) were manually added.

**Evaluation of the reliability of the 3D atomic coordinates**

To evaluate the validity of the experimentally determined 3D atomic coordinates, R-factors[4–6] between the experimental tilt series and forward-projections from the experimentally determined 3D atomic models were calculated. The averaged R-factors between the experimental tilt series and the forward-projections from the final 3D atomic models obtained from the two 3D tomograms [after applying each of the deep learning neural network trained by i) f.c.c. based atomic models and ii) amorphous based atomic models to the raw 3D tomogram] were 0.152 and 0.157, respectively. For a fair comparison, we ran the same test with the atomic models before manual corrections. The averaged R-factors between the experimental tilt series and the forward-projections from the final 3D atomic models obtained from the three 3D tomograms [i) before applying deep learning based neural network, ii) after applying deep learning neural network trained by f.c.c. based atomic models, and iii) after applying deep learning neural network trained by amorphous based atomic models] were 0.158, 0.151, and 0.156, respectively. It can be clearly seen that the deep learning based augmentation can better determine the atomic structure, as reflected in the lower R-factors. The result obtained from the deep learning neural network trained by the f.c.c. based atomic models showed the smallest R-factor, exhibiting the best consistency between the experimental tilt series images and the determined 3D atomic structure.

**Precision analysis for experimental data using multislice simulation**

To estimate the reliability of our final 3D atomic structure, we performed a precision analysis[3,4,6,8] using multislice simulation. A total of 21 projection images corresponding to the experimental tilt angles were obtained from the experimentally determined 3D atomic coordinates using the multislice simulation[14]. The multislice simulation was calculated with the parameters consistent with the experimental conditions: 300 keV electron energy, -775 nm $C_3$ aberration, 378 μm $C_5$ aberration, 25.1 mrad convergence semi-angle, and 40 mrad and 200 mrad detector inner and outer semi-angles. We also applied 8 frozen phonon configurations and 2 Å slice thickness for the simulation. A 3D tomogram was calculated from the multislice simulated tilt series using the GENFIRE algorithm[7]. To correct the imperfection of the 3D tomogram, the deep learning neural network[8] was subsequently applied. By running the same atomic-tracing and atom classification steps, two traced atomic models from the 3D tomograms before and after applying the deep learning neural network were obtained. The number of total uncommon atoms between experimentally determined atomic model and the two traced atomic models from the multislice simulated tomograms [i) before and ii) after applying the deep learning neural network] are 14 and 8, respectively.

We defined the error of atom identification as the ratio between the number of total uncommon atoms and the number of total atoms. The error of atom identification before applying the deep learning neural network was 0.4%, which was improved to 0.2% after applying the deep learning neural network. Root-mean-square deviation (i.e., precision) between the experimentally determined atomic model and the traced atomic models from the multislice images, before and after applying the deep learning neural network, are 16.3 pm and 9.9 pm, respectively (see Fig. S1).

**Estimation of disorder: 2D RMSD, BOO, and rotation angle**
The 2D root-mean-square deviation (2D RMSD) between an ideal 2D hexagonal lattice and each atomic layer perpendicular to [111] direction was calculated with the following procedures. First, an atom closest to the center in each atomic layer perpendicular to [111] direction was set as the origin of a 2D hexagonal lattice. Then, the nearest neighbor positions from the origin atom for an ideal 2D hexagonal lattice were calculated based on a nearest-neighbor bond length of 2.83 Å. If there is an atom within the distance of 0.7 Å from each of the calculated positions, the atom was assigned to the 2D hexagonal lattice site corresponding to the given nearest neighbor site of the origin. The search and assign processes were continued from the newly assigned atoms, and repeated until there is no newly assigned atoms to the lattice. Second, the obtained 2D hexagonal lattice was fitted to the corresponding experimentally measured atomic coordinates to determine the position, rotational orientation, and lattice parameter (bond length) of the 2D hexagonal lattice which minimizes the 2D RMSD between the experimental atomic coordinates and the fitted lattice. Then, using the fitted lattice parameters, the first step (search and assign process) and second step (fitting process) were repeated. This iterative process continued until the absolute sum of the differences between the new and old fitted 2D lattice vectors is smaller than $10^{-5}$ Å. The final 2D RMSDs between each atomic layer perpendicular to [111] direction and 2D fitted lattice are presented in Fig. S6b. Also, we separately calculated the 2D RMSD for surface atoms and core atoms in each layer perpendicular to [111] direction, respectively (see Fig. S6b). Surface atoms in each layer perpendicular to [111] direction were determined using alpha shape algorithm[15] with shrink factor 0.5, and the core atoms were defined as non-surface atoms. Additionally, the 2D rotation angles between the interface layer and the two adjacent layers were calculated based on the angular orientation of the fitted hexagonal lattice of each layer (see Fig. S6a).
We also calculated the overall relative rotation angle (along the [111] direction) between the top and bottom nanoclusters. For each of the top and bottom nanoclusters, we fitted a 3D f.c.c. lattice following the search and assign procedure described above (the only difference is that the search ran through the nearest neighbor positions of a 3D f.c.c. lattice rather than those of 2D hexagonal lattice). From the two fitted f.c.c. lattices, the relative rotation angle between the two nanoclusters along the [111] direction was determined to be 1.6° (see Fig. S6a).

The local bond orientation order (BOO) parameters for the experimental 3D atomic model were calculated using a method described in previous works[16,17]. The $Q_4$ and $Q_6$ order parameters (BOO parameters) were calculated based on a cutoff distance of 3.416 Å (the mean value of the nearest and next nearest neighbor distances of an f.c.c. structure with a lattice constant 3.912 Å). Figure S6c shows the $Q_4$ and $Q_6$ parameters calculated from the twin boundary layers, the interface layer, and the other layers. To quantify the disorder from the BOO parameters, the averaged 2D distance in the $Q_4$ v.s. $Q_6$ map between the BOO values of each atom in the final 3D atomic model and those of the ideal lattices (f.c.c. and h.c.p.) was calculated. The averaged distance between the BOO values of the upper (lower) twin boundary and that of an ideal h.c.p. lattice is 0.0324 (0.0318). The averaged distance between the BOO values of the interface and that of an ideal f.c.c. lattice is 0.0652. The averaged distance between the BOO values of the other atoms and that of an ideal f.c.c. lattice is 0.0160. These results indicate that the interface layer is much more disordered compared to the other parts of the Pt nanodumbbell, and the twin boundary layers adjacent to the interface layer also show more disorder, but in somewhat lesser degree compared to the interface.

**Strain map calculation**

Atomic displacements from an ideal f.c.c. lattice and corresponding 3D strain tensor were calculated from the experimentally determined 3D atomic positions following a method described in previous works[3]. In short, 3D fitted f.c.c. lattices corresponding to the top and bottom nanoclusters of the Pt nanodumbbell were obtained following the lattice fitting procedures described above (see the 'Estimation of disorder' part in Methods). The percentages of assigned atoms to the fitted f.c.c. lattices for the top and bottom nanoclusters were 99.0% and 99.4%, respectively. The RMSDs between the assigned atom positions and the fitted lattices were 38 and 32 pm for the top and bottom nanoclusters, respectively.

Atomic displacements were calculated by subtracting the fitted f.c.c. lattice positions from the assigned experimental atom positions. Atomic displacement field in a Cartesian grid was obtained by applying a Gaussian kernel of a standard deviation of 2.7 Å to the atomic displacements. After the process, 3D strain maps were computed by differentiating the atomic displacement fields.

**Surface local strain and ORR activity calculation**

First, local lattice constants for each atom were calculated using the following procedure.

i) We obtained a 3D globally fitted f.c.c. lattice following the 3D lattice fitting procedures described above (see the 'Estimation of disorder' part in Methods).

ii) Then, for each atom, we prepared a set of atoms which only contains atoms which corresponds to the nearest neighbor sites from the given atom, based on the globally assigned fitted lattice.

iii) For each atom, a local lattice was constructed as follows. For a given atom, we set the atom position as the origin of the local lattice. From the atom, the nearest neighbor f.c.c. lattice sites were calculated (initially it is calculated based on a nearest neighbor distance of 2.83 Å). If an atom in the prepared nearest neighbor atom set exists within the distance of 0.7 Å from each of the calculated lattice sites, the atom was assigned to the 3D local f.c.c. lattice site corresponding to the given nearest neighbor site. The search and assign process were repeated for all calculated nearest neighbor sites. Then, the obtained 3D local f.c.c. lattice was fitted to the assigned atomic coordinates to determine the position, rotational orientation, and lattice parameter (bond length) of the 3D local f.c.c. lattice which minimizes the RMSD between the assigned experimental atomic positions and the local fitted lattice.

iv) Using the newly fitted lattice parameters, the step iii) was repeated. This iterative process continued until the absolute sum of the differences between new and old fitted local f.c.c. lattice vectors is smaller than $10^{-5}$ Å.

v) The steps ii) - iv) were applied to all atoms in the nanoclusters.

From the obtained 3D local f.c.c. lattice for each atom, the local lattice constant was determined. Second, the local strain for each atom was calculated as $(a_{local} - a_{ref})/a_{ref}$, where $a_{local}$ is the obtained local lattice constant and $a_{ref}$ is the bulk Pt lattice constant 3.912 Å.

To define the {111} and {100} facets, we first set reference planes, the outmost atomic plane containing more than 20 atoms while being perpendicular to <111> and <100> directions, respectively. Then, each facet was defined as a set of atoms either in a corresponding reference plane or those outside the reference plane toward the facet normal direction. The surface local strains of {111} and {100} facets provided in Fig. 4a-f were mapped on a 3D Cartesian grid by applying a Gaussian kernel of a standard deviation of 2.7 Å to the calculated local strains of the facet atoms.

The ORR activity was directly calculated from the local strain using the relation between the local strain and ORR activity described in Fig. S7c. Also, the ORR activity for the {111} and {100} facets in Fig. 4g-l was similarly mapped on a 3D grid by applying a Gaussian kernel of a standard deviation of 2.7 Å to the calculated ORR.

## 2. Supplementary figures

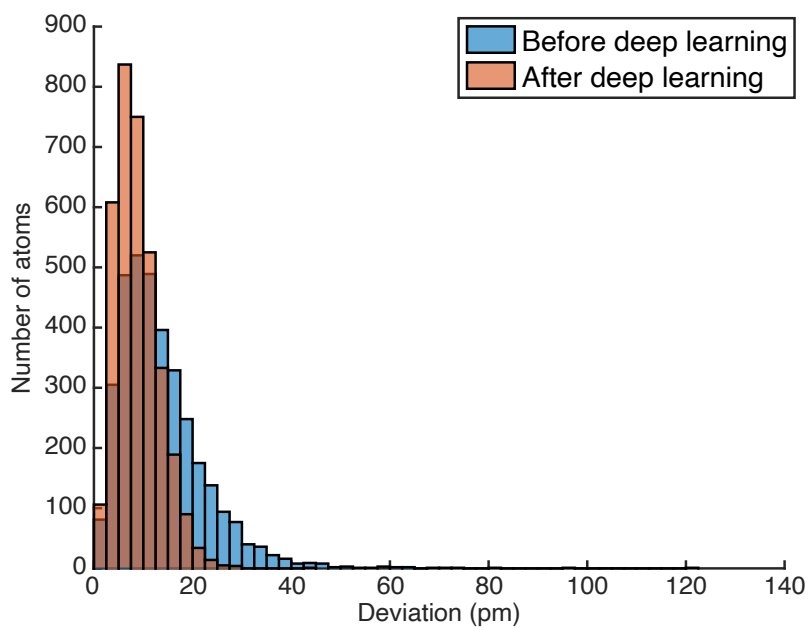

**Figure S1. Histogram of the deviation of atom positions between the experimentally measured atomic model of the Pt nanodumbbell and that obtained from the 3D tomograms from multislice simulated images, before and after applying the deep learning neural network.** The root-mean-square deviation (distance) between the experimental atomic model and that from the simulated 3D tomograms before applying the deep learning neural network was 16.3 pm, which was improved to 9.9 pm after applying the deep learning neural network.

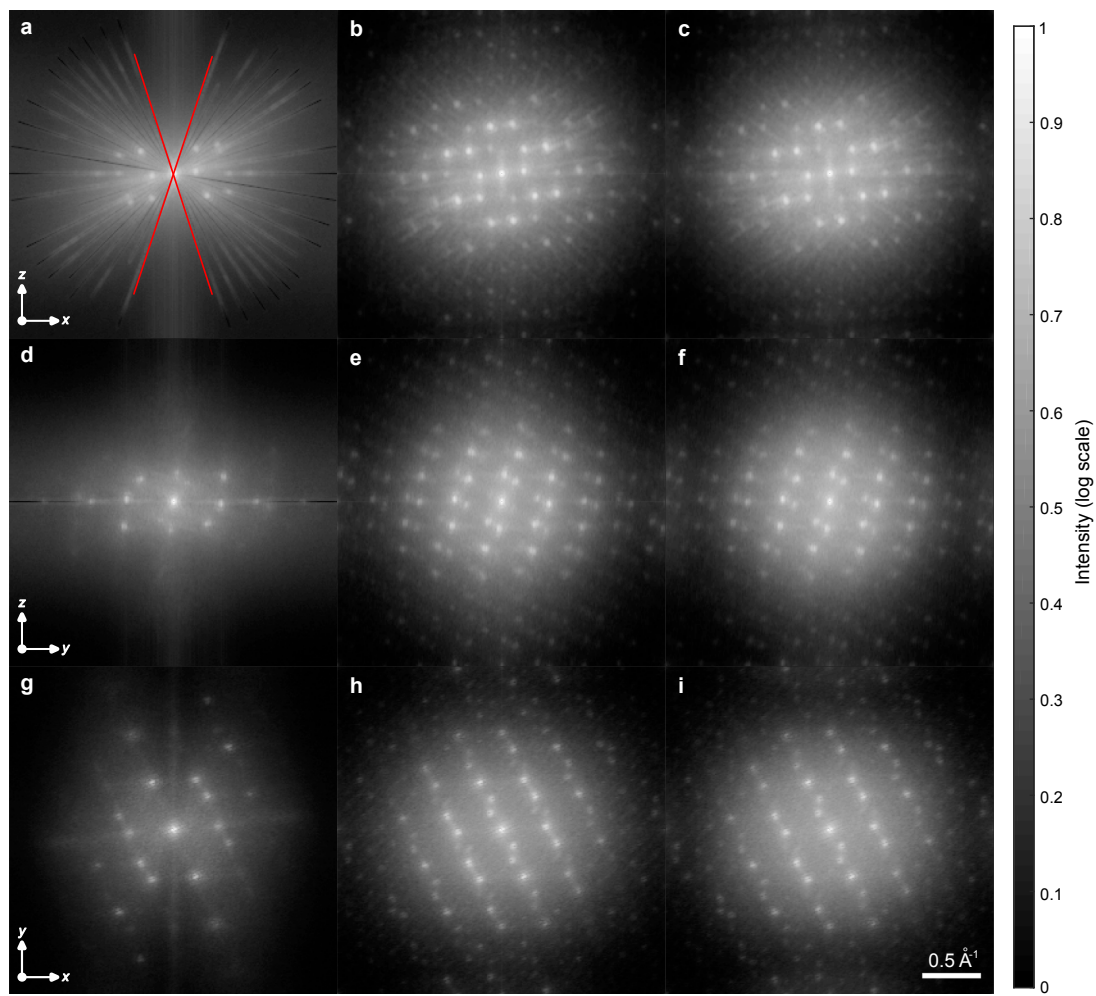

**Figure S2. Projected 3D Fourier intensities of experimental tomograms. a-c**. Projections of 3D Fourier intensities along the *y*-direction for the raw experimental tomogram of the Pt nanodumbbell (**a**), for the tomograms obtained by applying the deep learning augmentations trained by f.c.c. based models (**b**), and by amorphous based models (**c**), respectively. **d-f**. Projections similar to (**a-c**) along the *x*-direction. **g-i**. Projections similar to (**a-c**) along the *z*-direction. The red lines in (**a**) are the guidelines to visualize the missing wedge. It can be clearly seen that the missing wedge information is properly retrieved by the deep learning augmentations.

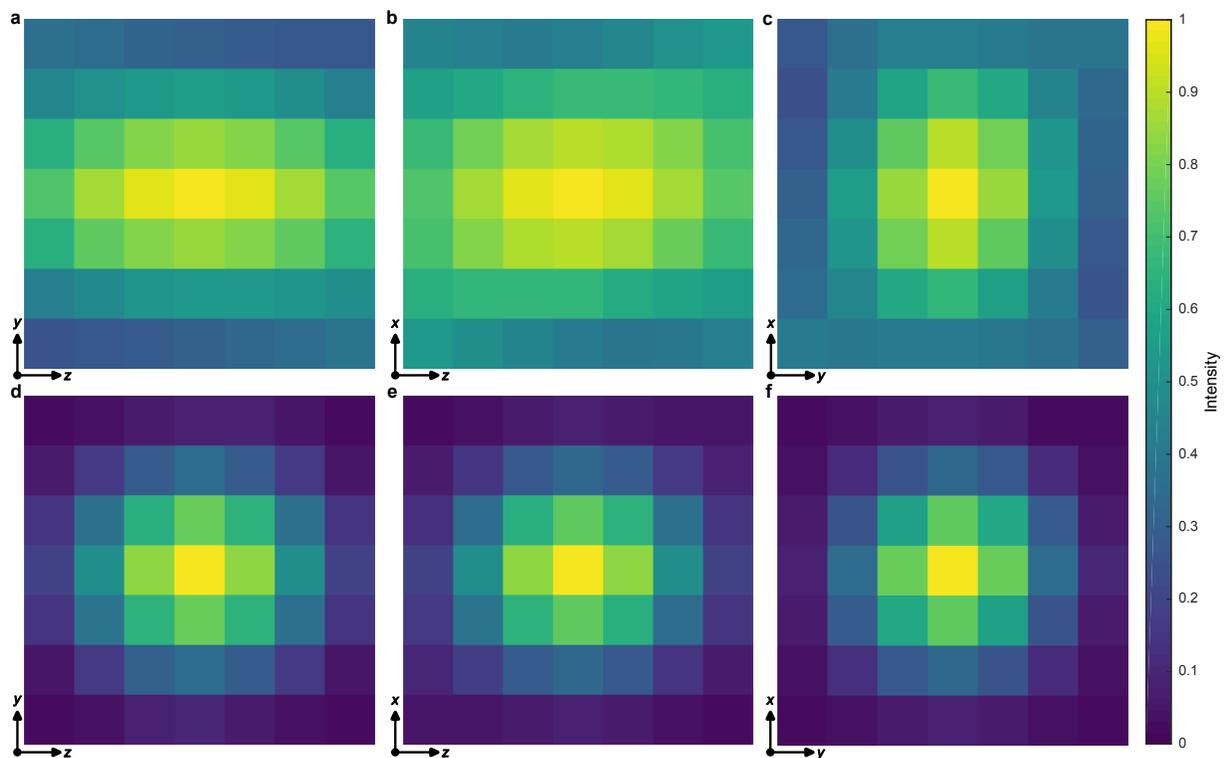

**Figure S3. Averaged 3D atom profiles obtained from the experimental data before and after applying the deep learning neural network. a-c**. Averaged intensity profiles (central slices) of all traced atoms from the raw tomogram, sliced perpendicular to the *x*-direction (**a**), the *y*-direction (**b**), and the *z*-direction (**c**). **d-f**. Averaged intensity profiles (central slices) of all traced atoms from the deep learning augmented tomogram, sliced perpendicular to the *x*-direction (**d**), the *y*-direction (**e**), and the *z*-direction (**f**). The pixel size is 0.358 Å.

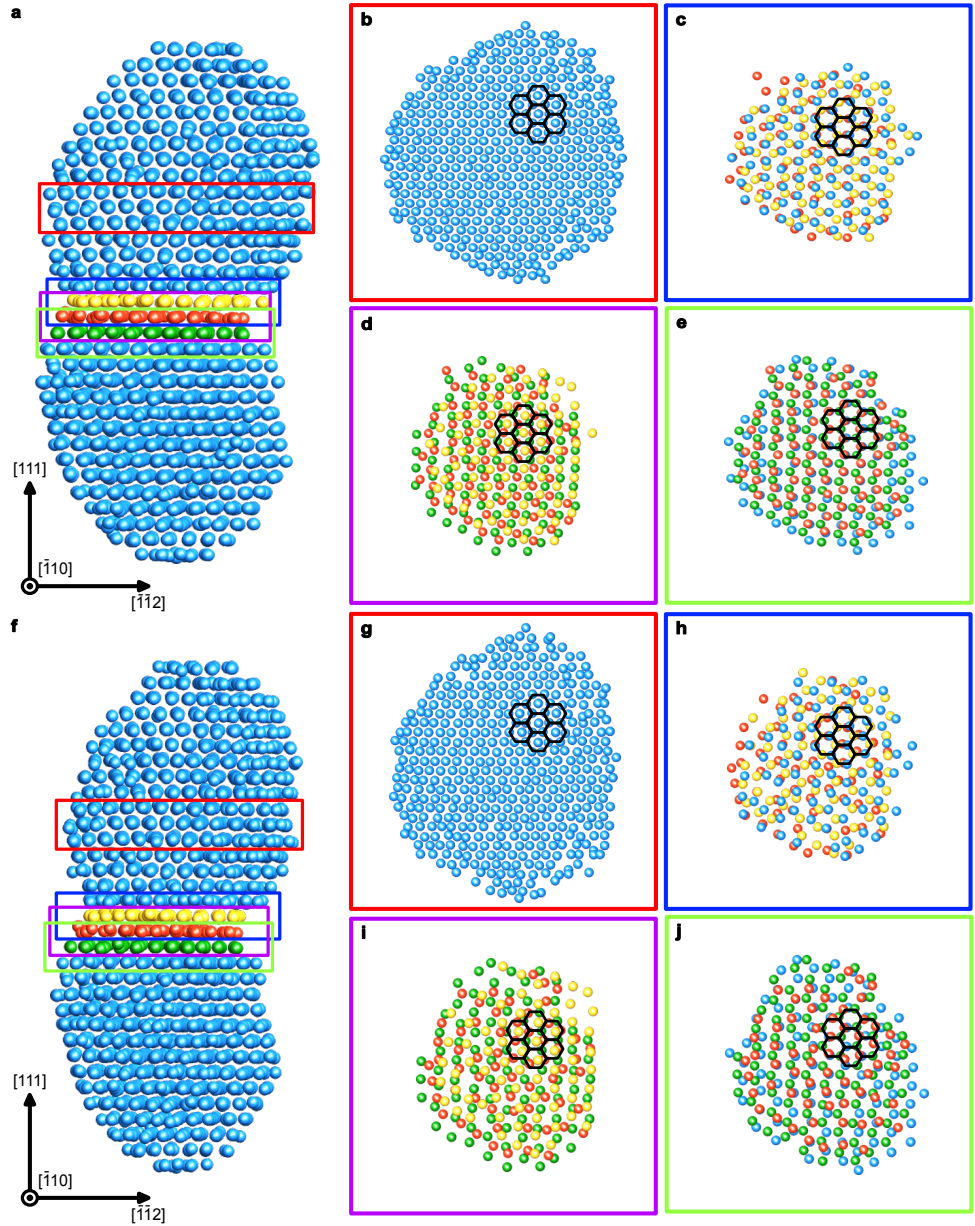

**Figure S4. Overall 3D structures, twin boundaries, and the interface structures of the Pt nanodumbbell obtained from the tomograms after applying the deep learning neural networks trained by different atomic models. a**. The 3D atomic structure obtained from the 3D tomogram after applying the deep learning neural network trained by f.c.c. based atomic models. The 3D atomic structure contains twin boundaries at the interface between the two nanoclusters. **b**. Three atomic layers perpendicular to [111] direction, deep inside the top nanocluster [the atomic layers in the red box in (**a**)], viewed along the [111] direction. **c**. Similar plot with (**b**), for the three atomic layers near the interface [the layers in the blue box in (**a**)]. **d**. Similar plot with (**b**), for the three atomic layers at the interface [the layers in the purple box in (**a**)]. **e**. Similar plot with (**b**), for the three atomic layers near the interface [the layers in the green box in (**a**)]. **f**. The 3D atomic structure obtained from the 3D tomogram after applying deep learning neural network trained by amorphous based atomic models. **g-j**. Similar plot with (**b**)-(**e**), for the atomic structure given in (**f**). The colors of the atoms follow the color convention in Fig. 2. The black hexagons in (**b-e, g-j**) are guidelines to highlight the f.c.c. and h.c.p. stacking orders.

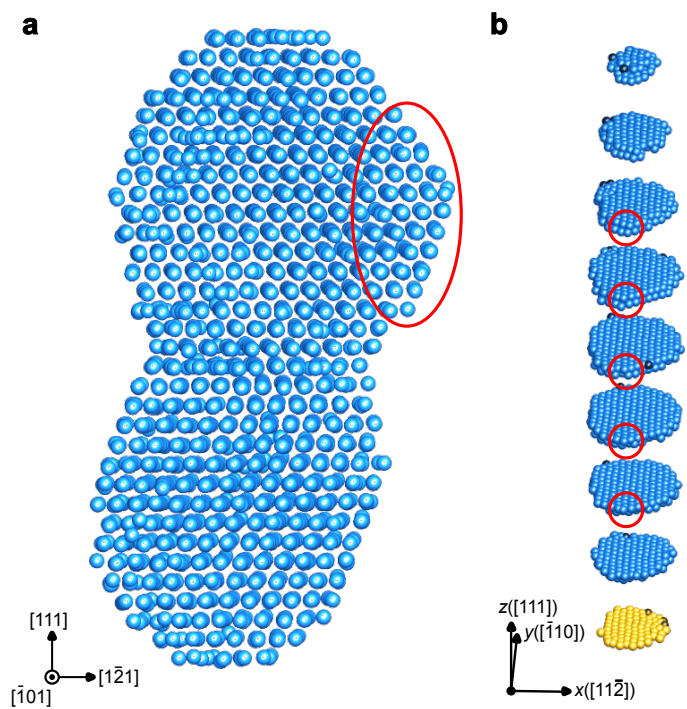

**Figure S5. The protruded region. a**. The overall 3D atomic structure of the Pt nanodumbbell. **b**. The 3D atomic structure of the Pt nanodumbbell sliced along the [111] direction for the top half of the Pt nanodumbbell. The red ellipses highlight the protruded region.

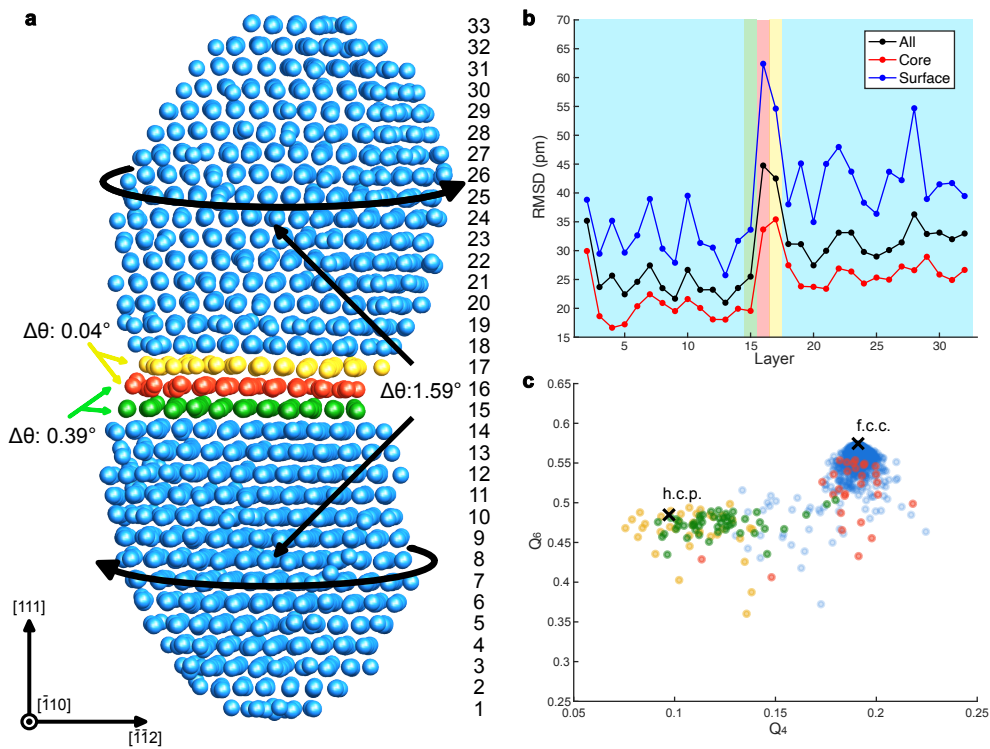

**Figure S6. Overall atomic structure, 2D root-mean-square deviation, and bond orientation order. a**. The atomic structure of the Pt nanodumbbell. The colors of the atoms follow the color convention in Fig. 2. The relative rotation angle along the [111] direction between the interfacial layer (the red atomic layer) and upper twin boundary (the yellow atomic layer) is 0.04°. Similarly, the rotated angle along [111] direction between the interfacial layer (the red atomic layer) and lower twin boundary (the green atomic layer) is 0.39°. The overall relative rotation angle between top and bottom nanoclusters was calculated to be 1.59°. **b**. The 2D RMSD for each atomic layer [each layer is assigned with a number as given in (**a**)] along [111] direction. **c**. The $Q_4$ and $Q_6$ values of bond orientation order for all traced Pt atoms. Cross marks represent the $Q_4$ and $Q_6$ values for ideal f.c.c. and h.c.p. structures, respectively. The colors of dots follow the color convention in (**a**).

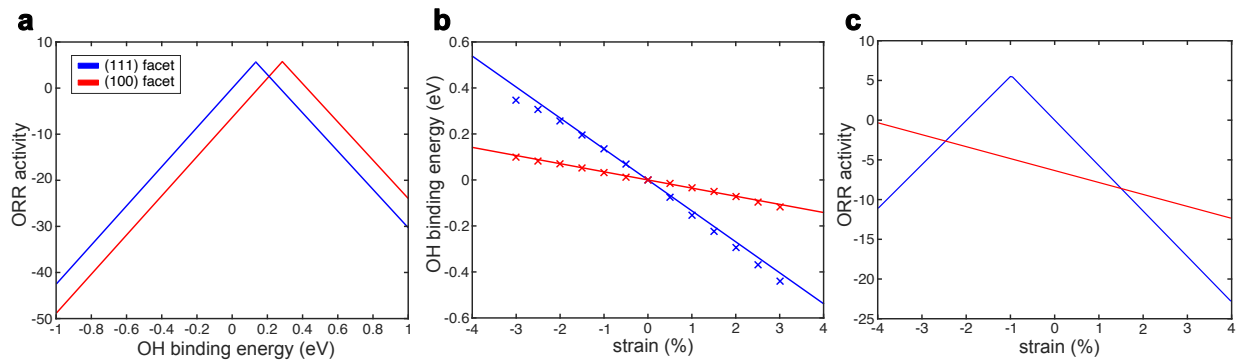

**Figure S7. The relation between ORR activity, OH binding energy, and strain. a**. Volcano-shaped relation between ORR activity and OH binding energy for {111} and {100} facets[18,19]. The OH binding energy in the *x* axis represents the relative change from that of an ideal Pt (111) surface. **b**. The relation between the OH binding energy and strain for both facets were calculated from density function theory (cross marks)[20], and the lines represent the linear relation obtained by a linear fitting. **c**. The final relation between ORR activity and strain for both facets were obtained by combining the relations in (**a**) and (**b**). The ORR activity is represented as $\log(j/j_{Pt,(111)})$, where $j$ is the current density.

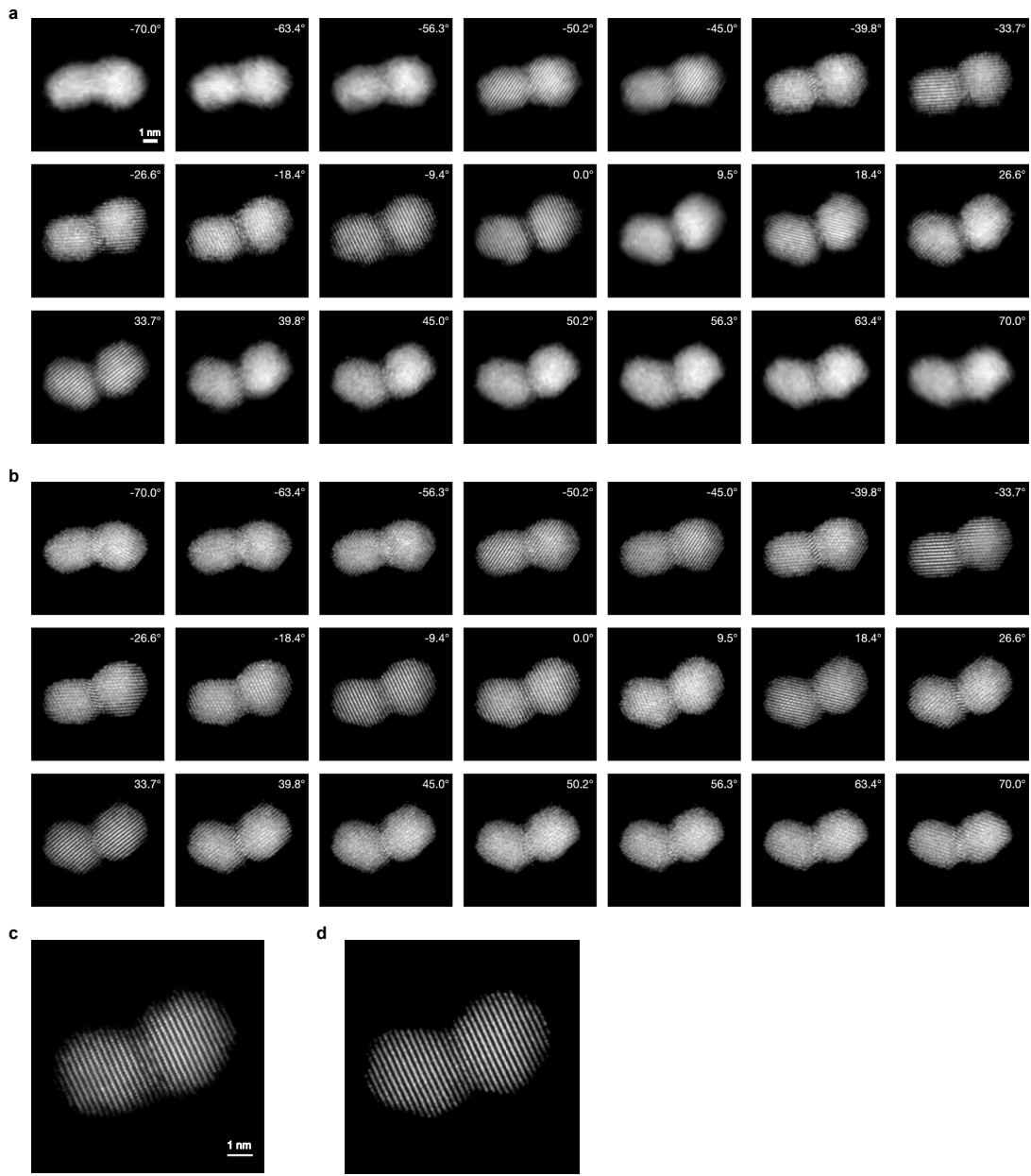

**Figure S8. Experimental tilt series and forward-projected tilt series from the final atomic model. a**. The post-processed tilt series images acquired from the ADF-STEM experiment (see Methods). The angles denoted at the top right corner of each image represent the $\theta$ tilt angles. **b**. Forward-projected tilt series from the final 3D atomic model along the tilt angles. Electron scattering factors of Pt atoms and Gaussian broadening (broadening width: 0.45 Å) were considered for the forward projections. Averaged R-factor[4,5] between the tilt series images in (**a**) and (**b**) is 0.151. **c**. The experimental zero-degree projection acquired right after the tilt series acquisition. **d**. The forward projection of the final 3D atomic model along the angle which gives the best consistency with the post-experiment zero-degree projection (**c**). The best consistency angle was determined to be $\psi : -0.4°, \theta : 2.6°, \varphi : -3.8°$.

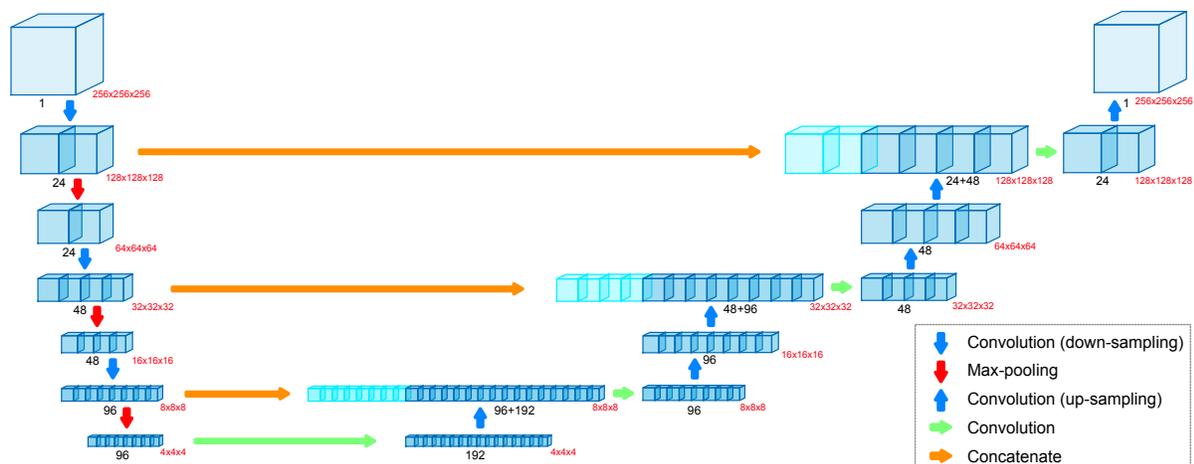

**Figure S9. The structure of the deep learning augmentation.** The framework of the deep learning neural network is based on a 3D-unet[21]. The set of boxes represents the 3D feature map. The black numbers below the 3D feature maps represent the number of channels for each 3D feature map. The red numbers next to the 3D feature maps represent the volume size of the 3D feature maps.

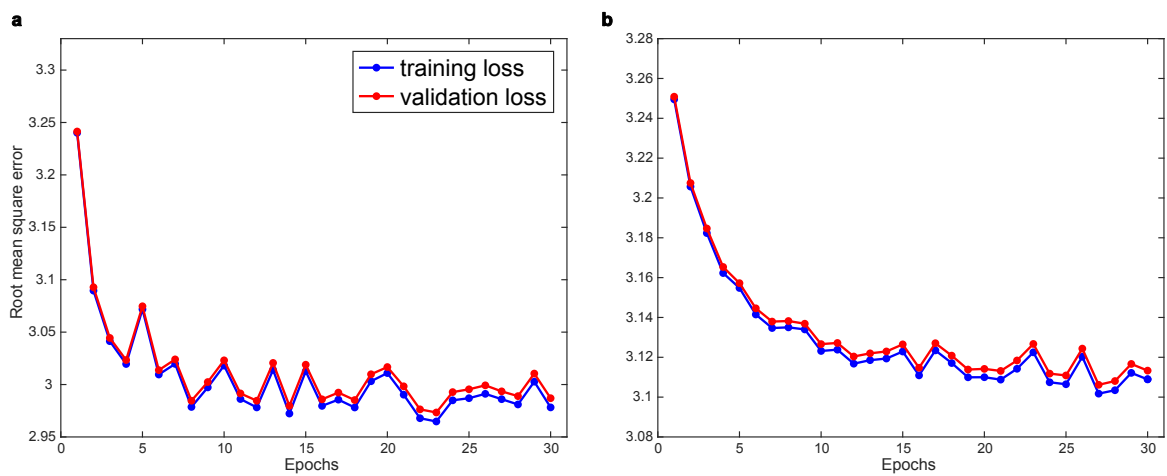

**Figure S10. Learning curves of the deep learning neural networks for two different training datasets.** The learning curves for the deep learning neural networks trained by (**a**) the f.c.c. atomic models, (**b**) the amorphous atomic models. The blue and red dots and lines represent the losses from training datasets and validation datasets, respectively.